\pdfoutput = 1
\documentclass[
	aps,
	prl,
	reprint,
	nofootinbib,
	nobibnotes,
	superscriptaddress,
	preprintnumbers,
	groupedaddress
]{revtex4-1}

\usepackage[utf8]{inputenc}
\usepackage[T1]{fontenc}
\usepackage{lmodern}
\usepackage{textcomp}
\usepackage{microtype}

\usepackage[abbreviations, british]{foreign}

\newcommand{\modified}[1]{\textcolor{black}{#1}}

\usepackage{upgreek}
\usepackage{mathtools}
\usepackage{mathrsfs}
\usepackage{fixmath}
\usepackage{alphabeta}
\usepackage{bm}
\usepackage{bbm}
\usepackage{physics}
\usepackage{units}
\usepackage{slashed}
\usepackage{isotope}

\usepackage[normalem]{ulem}

\usepackage{ragged2e}
\usepackage[inline]{enumitem}
\setlist{noitemsep}
\newlist{inlinelist}{enumerate*}{1}
\setlist*[inlinelist,1]{itemjoin={,\ }, itemjoin*={, and\ }, after=.}

\usepackage[dvipsnames]{xcolor}

\usepackage[subrefformat = parens]{subcaption}

\usepackage{threeparttable}
\DeclareCaptionFormat{revtex}{#1#2\justifying{#3}}
\usepackage{booktabs}
\usepackage{dcolumn}
\usepackage{graphicx}
\usepackage{amssymb}
\usepackage[normalem]{ulem}

\graphicspath{{./figures/}}
\providecommand{\tikzsetnextfilename}[1]{}

\newcommand{\rhoend}{\rho_{\rm end}}
\newcommand{\rhoreh}{\rho_{\rm re}}

\newcommand{\wrehbar}{\bar{w}_{\rm re}}

\newcommand{\Nreh}{N_{\rm re}}

\newcommand{\Vend}{\mathcal{V}_{\rm end}}
\newcommand{\g}{\mathsf{g}}

\newcommand{\V}{\mathcal{V}}

\newcommand{\SM}{\mathsf{a}}

\newcommand{\Treh}{T_{\rm re}}
\newcommand{\M}{M}
\newcommand{\vv}{\mathsf{v}}
\newcommand{\fluct}{\phi}
\newcommand{\GG}{\Gamma}
\newcommand{\Nk}{N_{\rm k}}
\newcommand{\epsilonvarM}{\frac{m_\phi}{M_{pl}}}
\newcommand{\epsilonvar}{\frac{m_\phi}{\varphi_{\rm end}}}
\newcommand{\nn}{{ n}}
\newcommand{\n}{ j}
\newcommand{\DD}{D}
\newcommand{\x}{x}
\newcommand{\sigmar}{\sigma_r}
\newcommand{\sigmans}{\sigma_{n_s}}
\newcommand{\nsbar}{\bar{n}_s}
\newcommand{\rbar}{\bar{r}}

\newcommand{\upalphatosigma}{\modified{\upsigma}}

\usepackage[hidelinks, pdfencoding = auto]{hyperref}
\usepackage[noabbrev, capitalize]{cleveref}
\crefname{enumi}{point}{points}
\makeatletter\DeclareRobustCommand{\labelcrefrange}[2]{\@crefrangenostar{labelcref}{#1}{#2}}\makeatother

\let\oldcite\cite\renewcommand\cite{\unskip~\oldcite}
\let\oldeqref\eqref\renewcommand\eqref{\unskip~\oldeqref}
\let\oldparagraph\paragraph\renewcommand{\paragraph}[1]{\oldparagraph{#1:}}

\begin{document}

\preprint{CP3-22-41}

\title{Connecting Cosmic Inflation to Particle Physics\\ with LiteBIRD, CMB-S4, EUCLID and SKA}

\author{Marco Drewes$^1$}
\email{marco.drewes@uclouvain.be}
\author{Lei Ming$^{2,1,3}$}
\email{minglei@mail.sysu.edu.cn (Corresponding author)}

\affiliation{$^1$Centre for Cosmology, Particle Physics and Phenomenology, Université catholique de Louvain, Louvain-la-Neuve B-1348, Belgium}

\affiliation{$^2$ School of Physics, Sun Yat-Sen University,
Guangzhou 510275, China}

\affiliation{$^3$School of Physics, Nanjing University, Nanjing, 210093, China}

\begin{abstract}
    \noindent We show that next generation Cosmic Microwave Background experiments will be capable of the first ever measurement of the inflaton coupling to other particles, opening a new window to probe the connection between cosmic inflation and particle physics.This sensitivity is based on the impact that the reheating phase after cosmic inflation has on the redshifting of cosmic perturbations. For our analysis we introduce a simple analytic method to estimate the sensitivity of future CMB observations to the reheating temperature and the inflaton coupling. Applying our method to LiteBIRD and CMB-S4 we find that,  within a given model of inflation, these missions have the potential to impose both an upper and a lower bound on the inflaton coupling. Further improvement can be achieved if CMB data is combined with optical and 21cm surveys. Our results demonstrate the potential of future observations to constrain microphysical parameters that can provide an important clue to understand how a given model of inflation may be embedded in a more fundamental theory of nature.
\end{abstract}

\maketitle

\paragraph{ Introduction}
The current concordance model of cosmology, known as $\Lambda$CDM model, can explain almost all properties of the observable universe at an astonishing accuracy with only a handful of free parameters  \cite{Planck:2018vyg,BICEP:2021xfz}.\footnote{See \cite{Abdalla:2022yfr} for a recent summary of tensions and anomalies in the $\Lambda$CDM model.} 
Leaving aside the composition of the Dark Matter (DM),  
the model is firmly based on the Standard Model (SM) of particle physics and the theory of General Relativity (GR),
implying that the most fundamental laws of nature that we know from earth  \cite{Workman:2022ynf} hold in the most distant regions of the observable universe.\footnote{The value of the cosmological constant also has no microscopic explanation in the SM, but it can be regarded as a free parameter in GR and quantum field theory, reducing this issue to a theoretical problem of naturalness (cf.~\cite{Giudice:2008bi}) rather than a failure to describe data.}

However, to date it is unknown what mechanism set the initial conditions for the hot big bang, including the initial overall geometry of the observable Universe and the temperature $\Treh$ of the primordial plasma at the onset of the radiation dominated epoch.\footnote{Another initial condition that cannot be explained within the SM but is not addressed in the present work is the initial matter-antimatter asymmetry (cf.~\cite{Canetti:2012zc}).} 
The former -- in particular overall homogeneity, isotropy, and spacial flatness reflected in the Cosmic Microwave Background (CMB) -- is amongst the most compelling mysteries of modern cosmology. 
\emph{Cosmic inflation} \cite{Starobinsky:1980te,Guth:1980zm,Linde:1981mu} offers an elegant solution for these problems and can in addition explain the observed correlations amongst the small perturbations in the CMB.  
However, very little is known about the mechanism that may have driven the exponential growth of the scale factor $a$. 
A wide range of theoretical models of inflation exist (see e.g.~\cite{Martin:2013tda} for a partial list), but the observational evidence is not conclusive enough to clearly single out one of them. Moreover, even less is known about 
the embedding of cosmic inflation into a more fundamental theory of nature and its connection to theories of particle physics beyond the SM. 
In the next decade the observational situation will change drastically. 
Upgrades at the South Pole Observatory \cite{Moncelsi:2020ppj}
and the Simmons Observatory \cite{SimonsObservatory:2018koc} aim at pushing the uncertainty in
the scalar-to-tensor ratio 
$r$ down to $\delta r \sim 3\times 10^{-3}$. 
In the 2030s JAXA's LiteBIRD satellite \cite{LiteBIRD:2022cnt}
and the ground-based CBM-S4 program \cite{CMB-S4:2020lpa} are expected to further reduce this to $\delta r < 10^{-3}$ for $ r = 0$.

In the present work we for the first time quantify the ability of these missions to probe the connection between inflation and particle physics. 
We utilise the impact of \emph{cosmic reheating}~\cite{Albrecht:1982mp,Dolgov:1989us,Traschen:1990sw,Shtanov:1994ce,Kofman:1994rk,Boyanovsky:1996sq,Kofman:1997yn} after inflation on the expansion history, i.e., the dissipative transfer of energy from the inflationary sector to other degrees of freedom that filled the universe with particles and determined $\Treh$. 
While the only known direct messenger from the reheating epoch would be gravitational waves (cf.~\cite{Caprini:2018mtu}),\footnote{The thermal graviton background \cite{Ghiglieri:2015nfa}  can in principle be used to probe $\Treh$ \cite{Ghiglieri:2020mhm,Ringwald:2020ist}, but this is very challenging in practice \cite{Drewes:2023oxg}. } 
it can be studied indirectly with
CMB observables \cite{Martin:2010kz,Adshead:2010mc,Easther:2011yq} due to the impact of the modified equation of state $w$ during reheating \cite{Lozanov:2016hid,Lozanov:2017hjm} on the post-inflationary expansion history. This has motivated studies in various models.\footnote{Examples include
natural inflation \cite{Munoz:2014eqa,Cook:2015vqa,Wu:2018vuj,Stein:2021uge},
power law \cite{Cai:2015soa,DiMarco:2018bnw,Maity:2018qhi,Maity:2019ltu,Antusch:2020iyq} and polynomial potentials \cite{Dai:2014jja,Cook:2015vqa,Domcke:2015iaa,Dalianis:2016wpu},
Starobinski inflation \cite{Cook:2015vqa},
Higgs inflation \cite{Gong:2015qha,Cook:2015vqa,Cai:2015soa},
hilltop type inflation \cite{Cook:2015vqa,Cai:2015soa},
axion inflation \cite{Cai:2015soa,Takahashi:2019qmh},
curvaton models \cite{Hardwick:2016whe},
$\alpha$-attractor inflation \cite{Ueno:2016dim,Nozari:2017rta,DiMarco:2017zek,Drewes:2017fmn,Maity:2018dgy,Rashidi:2018ois,German:2020cbw,Mishra:2021wkm,Ellis:2021kad}, 
tachyon inflation \cite{Nautiyal:2018lyq}, 
inflection point inflation \cite{Choi:2016eif},
fiber inflation \cite{Cabella:2017zsa},
K\"ahler moduli inflation  \cite{Kabir:2016kdh,Bhattacharya:2017ysa},
and other SUSY models \cite{Cai:2015soa,Dalianis:2018afb}.}
 Reheating is inherently sensitive to the inflaton couplings to other fields, i.e.,
microphysical parameters that connect inflation to particle physics \cite{Drewes:2015coa,Martin:2016iqo,Drewes:2019rxn}, as these interactions mediated the energy transfer. 
However, past studies have almost exclusively focused on $\Treh$, ignoring the possibility to constrain microphysical (particle physics) parameters. 
The fundamental limitations on the possibility
of constraining the microphysical coupling constant $\g$ associated with the interaction through which the universe was primarily reheated were laid out in \cite{Drewes:2019rxn}, where it was estimated that such a measurement may be within reach of next-generation instruments. 
However, neither there nor in any of the few phenomenological works addressing the relation to microphysical parameters  \cite{Ueno:2016dim,Drewes:2017fmn,Ellis:2021kad} a systematic study was performed to quantify the feasibility of such a measurement with realistic instrumental sensitivities. 
In the present work we introduce a simple analytic method to quantify the sensitivity of observations to $\g$ for given instrumental sensitivities to
\modified{the amplitude of the scalar perturbations in the CMB $A_s$, the spectral index $n_s$ and the tensor-to-scalar ratio $r$.}\footnote{\modified{Throughout we fix  $A_s=10^{-10}e^{3.043}$ \cite{Planck:2018vyg} and neglect the uncertainty $\sigma_{A_s}$; we checked in \cite{Drewes:2023bbs} that this does not affect the conclusions.}}
We apply this method to show for the first time that upcoming observations will be capable of the first ever measurement of both $\g$ and $\Treh$, where we define a measurement as the ability to impose both an upper and a lower bound on the respective quantity.
In this Letter we present the main results of our research, a more detailed analysis is presented in \cite{Drewes:2023bbs}.

\paragraph{Imprint of reheating in the CMB}
The primary goal of this work is to quantify constraints on 
$\g$ and $\Treh$ 
from current and future CMB data.
We consider inflationary models that can effectively be described by a single field $\Phi$ and assume that the effective single field description holds throughout both, inflation and the reheating epoch.\footnote{
See \cite{Renaux-Petel:2015mga,Passaglia:2021upk,Drewes:2019rxn} and references therein for a discussion of effects that limit the validity of this assumption during reheating.
} 
Defining $\varphi=\langle\Phi\rangle$ as the quantum statistical expectation value of $\Phi=\varphi + \fluct$ with fluctuations $\fluct$ and $\varphi_{\rm end}$ as its value 
at the end of inflation, the energy density at the end of inflation $\rhoend\simeq \frac{4}{3}\V(\varphi_{\rm end}) \equiv \frac{4}{3}\Vend$ and the spectrum of primordial perturbations are fixed by the effective potential $\V(\varphi)$.\footnote{Throughout we work at leading order in the slow roll parameters $\epsilon$ and $\eta$ defined after \eqref{nANDr}; we confirmed in \cite{Drewes:2023bbs} that this approximation does not affect our conclusions.}
Assuming a standard cosmic history after reheating (and leaving aside foreground effects), 
the observable spectrum of CMB perturbations can be predicted from $\V(\varphi)$ once the expansion history during reheating is known. The latter requires knowledge of the duration of the reheating epoch $\Nreh$ in terms of expansion $e$-folds $N$ 
(defined as the logarithm of the scale factor growth)
and the average equation of state during reheating $\wrehbar = \frac{1}{N_{\rm re}}\int_0^{N_{\rm re}} w(N) dN$. 
Since the total energy density of the universe $\rho$ is still dominated by the energy density  $\rho_\varphi$ of $\varphi$ during reheating, $\wrehbar$ is determined by specification of $\V(\varphi)$, and $\Nreh$ is the only relevant quantity that is not fixed by the choice of $\V(\varphi)$, i.e., is sensitive to $\g$. 
Within a given model of inflation one can obtain information about $\Nreh$ by comparing the observed CMB spectrum to the model's prediction. 
Using the general redshifting relation $\rho \propto \exp(- 3N(1 + w))$,
this can then be translated into a constraint on the energy density at the end of reheating $\rhoreh= \rhoend\exp(- 3\Nreh(1 + \wrehbar))$ \cite{Lidsey:1995np}, often expressed in terms of an effective reheating temperature defined as $\frac{\pi^2 g_*}{30}T_{\rm re}^4 \equiv \rhoreh$ \modified{with $g_*$ the effective number of relativistic degrees of freedom},
\begin{equation}\label{Tre}
	\Treh=\exp\left[-\frac{3(1+\bar{w}_{\rm re})}{4}N_{\rm re}\right]\left(\frac{40 \ \modified{\Vend} }{g_*\pi^2}\right)^{1/4}.
\end{equation}
In order to further translate knowledge on $\Nreh$ into knowledge on microphysical parameters, we utilise the fact that reheating ends when $H=\GG$, where $\GG$ is an effective dissipation rate for $\varphi$ and $H=\dot{a}/a$ is the Hubble rate. 
Together with the Friedmann equation $H^2=\rho/(3 M_{pl}^2)$
this yields \cite{Drewes:2015coa,Drewes:2017fmn}
 \begin{eqnarray}
\GG|_{\GG=H}=\frac{1}{ M_{pl}}\left(\frac{\rhoend}{3}\right)^{1/2} e^{-3(1+\wrehbar) \Nreh/2}\, \label{GammaConstraint}
\end{eqnarray}
with $M_{pl}\modified{=2.435\times 10^{18}~{\rm GeV}}$ the reduced Planck mass. 
The RHS of \eqref{Tre} 
and \eqref{GammaConstraint} only contain quantities that are either calculable for given $\V(\varphi)$ or can be obtained from CMB observations; we summarise the relations to CMB observables in the appendix.
Meanwhile $\GG$ on the LHS depends on microphysical parameters of the particle physics model in which $\V(\varphi)$ is realised.


\paragraph{Measuring the inflaton coupling in the CMB}
We classify microphysical parameters in three categories.
A \emph{model of inflation} is defined by specifying the effective potential $\V(\varphi)$. 
Ignoring quantum corrections to the $\varphi$-trajectory, this is equivalent to fixing the set of coefficients $\{\vv_i\}$ of all operators in the action that can be constructed from $\Phi$ alone, e.g., by Taylor expanding the inflaton potential around its minimum as $ \V(\varphi) = \sum_\n
    \frac{\vv_{\n}}{\n !} \frac{\varphi^\n}{\Lambda^{\n-4}}  = \frac{1}{2}m_\phi^2\varphi^2+\frac{g_\phi}{3!}\varphi^3+\frac{\lambda_\phi}{4!}\varphi^4 + \ldots$.
The set of \emph{inflaton couplings} $\{\g_i\}$ comprises coupling constants (or Wilson coefficients) associated with operators that are constructed from $\Phi$ and other fields.
A complete particle physics model contains a much larger set of parameters than the combined set $\{\vv_i\} \cup \{\g_i\}$, including  the masses of the particles produced during reheating as well as their interactions amongst each other and with all other fields. We refer to the set of all parameters in the action that are not contained in $\{\vv_i\} \cup \{\g_i\}$  as $\{\SM_i\}$. This set, e.g., contains the parameters of the SM.

$\GG$ in \eqref{GammaConstraint} necessarily depends on the $\{\g_i\}$ and $\{\vv_i\}$. For instance, for reheating through elementary particle decays, one typically finds 
$\GG = \g^2 m_\phi /{c}$, with $\g\in \{\g_i\}$ a coupling constant,  $m_\phi \in \{\vv_i\}$ the inflaton mass, and $c$ a numerical factor. However, in general feedback effects 
from produced particles on the ongoing reheating process
introduce a dependence of $\GG$ on a large sub-set of $\{\SM_i\}$, making it impossible to determine $\g$ from the CMB in a model-independent way, i.e., without having to specify the details of the underlying particle physics model and the values of the parameters $\{\SM_i\}$.\footnote{This tends to happen due to resonant particle production during the so-called preheating phase (c.f.~\cite{Amin:2014eta} for a review), but can in principle also occur due to thermal feedback during perturbative reheating \cite{Kolb:2003ke,Mukaida:2012qn,Mukaida:2012bz,Drewes:2013iaa,Mukaida:2013xxa,Drewes:2014pfa,Drewes:2015coa,Drewes:2017fmn,Drewes:2019rxn,Co:2020xaf,Garcia:2020wiy}.} 
The conditions under which $\g$ can be constrained model-independently have been studied in detail
in \cite{Drewes:2019rxn}, a conservative estimate is
\begin{eqnarray}
|\g| \ll\left(\frac{m_\phi}{\varphi_{\rm end}}\right)^{\n-\frac{1}{2}}
{\rm min}\left(
\sqrt{\epsilonvarM}
,
\sqrt{\epsilonvar}
\right)
\left(\frac{m_\phi}{\Lambda}\right)^{4-\DD} \ \label{GeneralScaling}\\ 
|\vv_i| \ll \left(\frac{m_\phi}{\varphi_{\rm end}}\right)^{\n-\frac{5
}{2}}
{\rm min}\left(
\sqrt{\epsilonvarM}
,
\sqrt{\epsilonvar}
\right)
\left(\frac{m_\phi}{\Lambda}\right)^{4-\n}\label{GeneralScalingSelf}
\end{eqnarray}
 with $\DD$ the mass dimension of the interaction term under consideration, $\n$ the power at which $\Phi$ appears in that operator, and $\Lambda$ a scale that can be identified with $m_\phi$ for $D\leq4$ and represents a UV cutoff of the effective theory for $D>4$. 
 The conditions \eqref{GeneralScaling} and \eqref{GeneralScalingSelf} ensure that the production of particles proceeds slow enough that redshifting spreads their momenta over a sufficiently broad phase space volume for the occupation numbers in each mode to remain low enough to avoid sizeable feedback effects, such as a parametric resonance. 
The condition \eqref{GeneralScalingSelf} practically restricts the possibility to constrain $\g$ model-independently to scenarios where the $\varphi$-oscillations occur in a mildly non-linear regime. 

\begin{figure}[ht!]
	\centering
	\begin{subfigure}[b]{\columnwidth}
		\includegraphics[width = 0.9\linewidth]{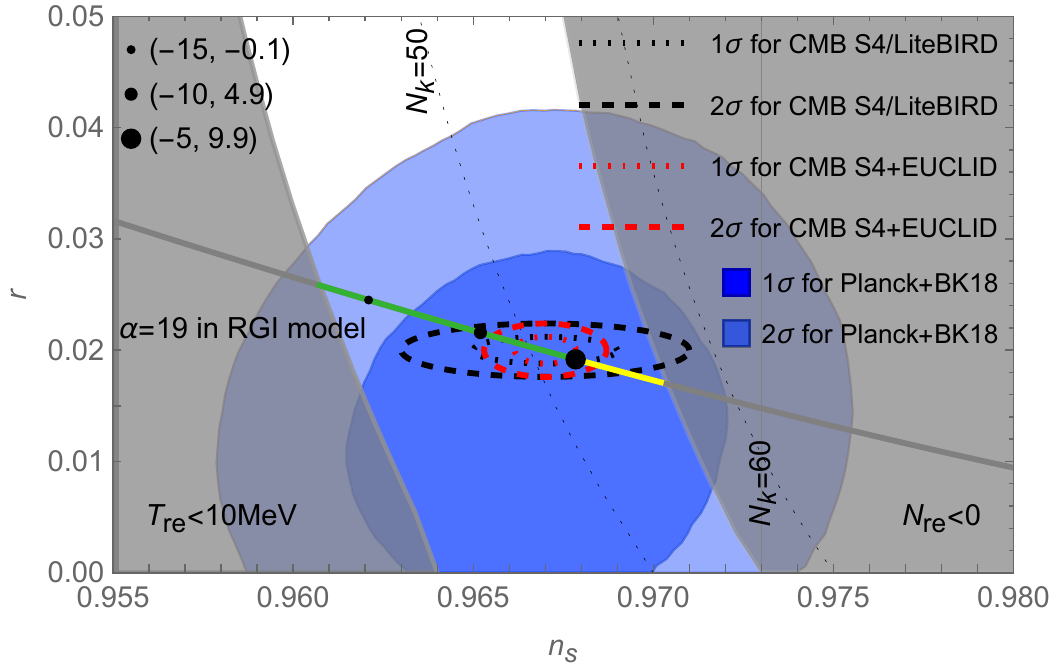}
	\end{subfigure}
	\\
	\begin{subfigure}[b]{\columnwidth}
		\includegraphics[width = 0.9\linewidth]{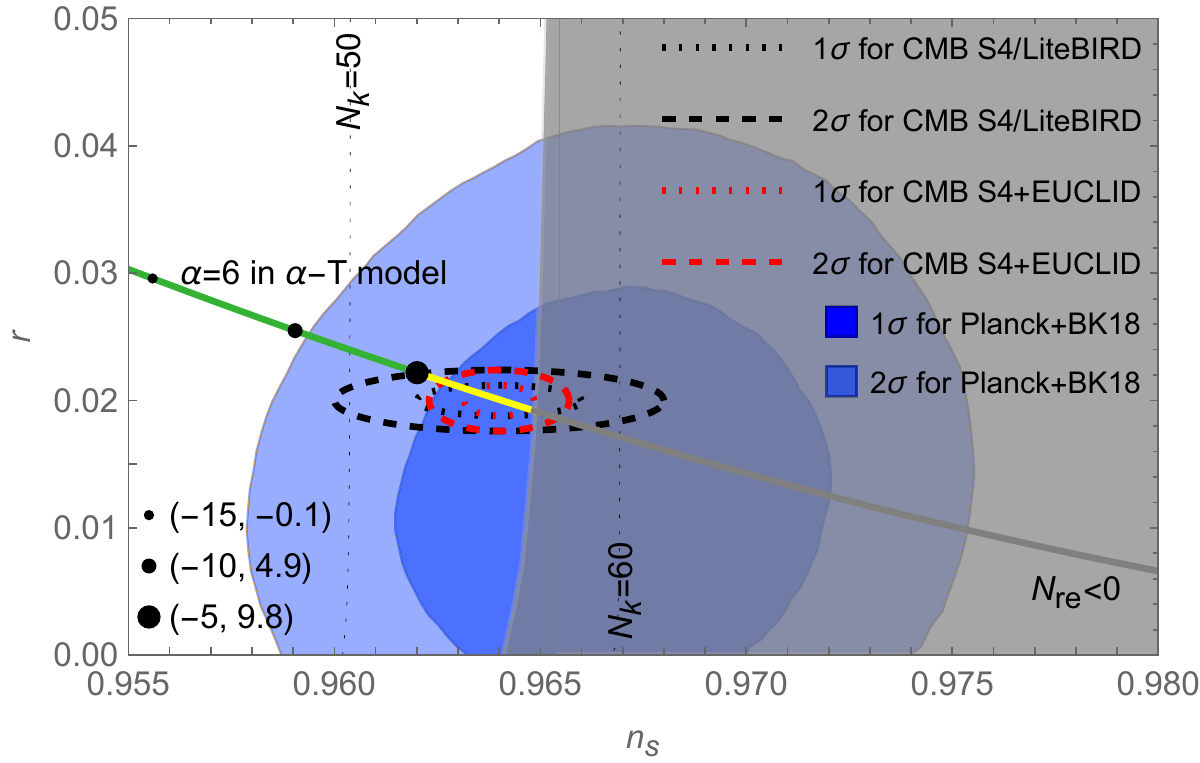}
	\end{subfigure}
	\caption{The diagonal line represents values of $n_s$ and $r$ predicted by the 
	RGI model (upper panel) and
	$\alpha$-T model (lower panel) for fixed $\alpha$, with the inflaton coupling varying along the curve.
 The black discs indicate the predictions for specific values of a Yukawa coupling $y$, with $(\log_{10}y,\log_{10}\Treh/{\rm GeV})$ given in the legend. 
 Conditions \eqref{GeneralScaling} and \eqref{GeneralScalingSelf} are fulfilled in the green part,
	the gray parts are ruled out by the conditions $\Nreh>0$ and $\Treh < T_{\rm BBN}$. Ellipses indicate current constraints and future sensitivities to $n_s$ and $r$, $\Nk$  is the  number of $e$-folds between the horizon crossing of a perturbation with wave number $k$ and the end of inflation \eqref{Nk}.
	\label{Figure1}
	}
	\label{results}
\end{figure}

\paragraph{Application to specific models}
In the following we apply the previous considerations to two models of inflation, namely radion gauge inflation (RGI)~\cite{Fairbairn:2003yx,Martin:2013nzq} and $\alpha$-attractor T-models ($\alpha$-T)~\cite{Kallosh:2013maa,Kallosh:2013hoa,Carrasco:2015pla,Carrasco:2015rva}, with the potentials 
\begin{eqnarray}
 {\rm RGI: \ }  \ \V(\varphi)&=&M^4\frac{(\varphi/M_{pl})^2}{\alpha+(\varphi/M_{pl})^2}   
 \label{RGI V} \\
 \alpha{\rm-T: \ } \ \V(\varphi)&=&M^4{\rm tanh}^{2\nn}
 \left(\frac{\varphi}{\sqrt{6\alpha}M_{pl}}\right) \ 
 \label{alpha V}  
\end{eqnarray}
The scale $M$ can be expressed in terms of other parameters with the help of \eqref{H_k},
\begin{eqnarray}
   {\rm RGI: \ }  \  M &=&
     M_{pl}
     \left(
     \frac{3\pi^2}{2} r A_s
     \left(1 + \alpha \frac{M^2_{pl}}{\varphi_k^2}\right)
     \right)^{1/4},
    \label{M2}\\
    \alpha{\rm-T: \ } \ M &=&
     M_{pl}
     \left(
     \frac{3\pi^2}{2}
     A_s r
     \right)^{1/4}
     \tanh^{-\frac{\nn}{2}}\left(
     \frac{\varphi_k}{\sqrt{6\alpha}M_{pl}}
     \right), \quad \
    \label{M3}
\end{eqnarray}
and condition \eqref{GeneralScalingSelf} implies $\nn=1$. Within these families of models 
\eqref{TakaTukaUltras} 
implies the relations
    $\alpha=\frac{432r^2}{(8(1-n_s)+r)^2(4(1-n_s)-r)}$
    and
    $\alpha=\frac{4r}{3(1-n_s)(4(1-n_s)-r)}$ 
for the RGI and $\alpha$-T models, respectively.
This defines a line in the $n_s$-$r$ plane, the position along which is given by $\Nreh$ (and hence $\g$), cf.~Fig.~\ref{Figure1}.
Condition 
\eqref{GeneralScalingSelf} implies $\alpha>2.4$ in \eqref{RGI V} and $\alpha>1/4$ in \eqref{alpha V}.
For our analysis we pick $\alpha=19$ in \eqref{RGI V} and $\alpha=6$ in \eqref{alpha V}.
When conditions \eqref{GeneralScaling} and \eqref{GeneralScalingSelf} are fulfilled we may parameterise  $\GG = \g^2 m_\phi /{c}$ \cite{Drewes:2019rxn} with $(\g,c)=(g/m_\phi,8\pi)$ for a scalar coupling $g\Phi\chi^2$ \cite{Boyanovsky:2004dj}, $(\g,c)=(y,  \modified{8\pi})$ for a Yukawa coupling $y\Phi\bar{\psi}\psi$ \cite{Drewes:2013iaa},
and $(\g,c)=(\upalphatosigma \modified{m_\phi/\Lambda},4\pi)$ for an axion-like coupling $\frac{\upalphatosigma}{\Lambda}\Phi F_{\mu\nu}\tilde{F}^{\mu\nu}$ \cite{Carenza:2019vzg}, where we neglected the produced particles' rest masses. 
We shall assume a Yukawa coupling $y$ in the following, bounds on other interactions can be obtained by simple rescaling according to $c$ \cite{Drewes:2023bbs}. 

\begin{figure}[ht!]
	\centering
	\begin{subfigure}[b]{\columnwidth}
		\includegraphics[width = 0.9\linewidth]{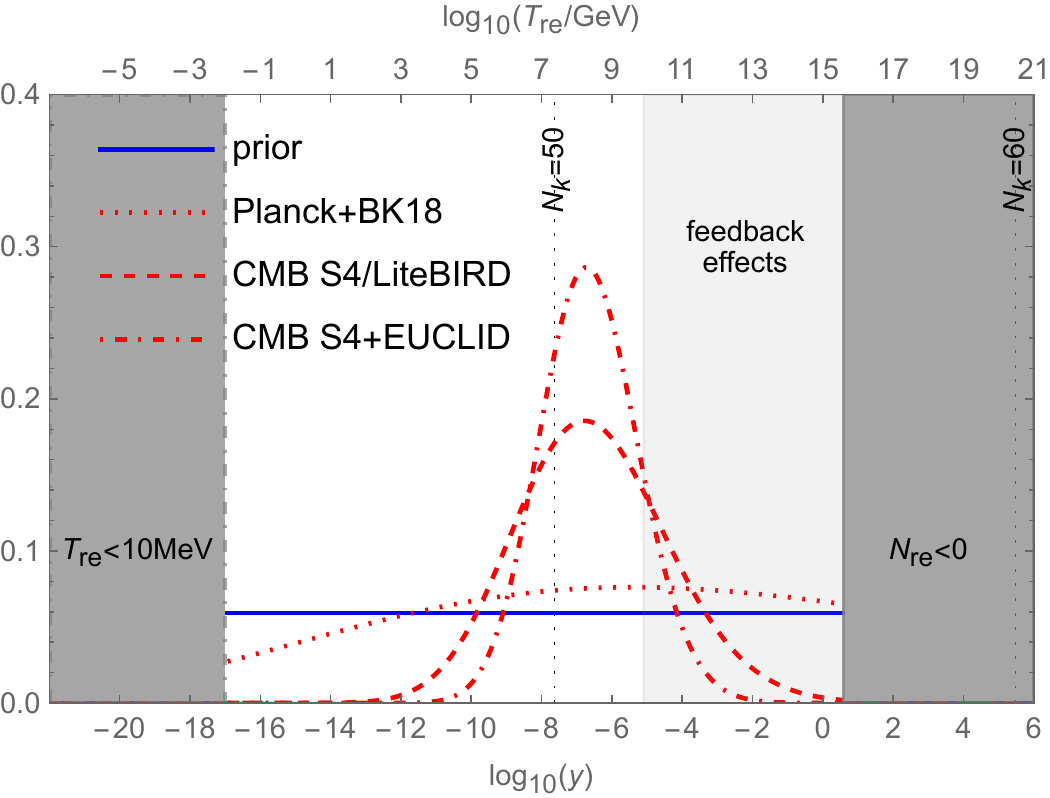}
	\end{subfigure}
	\\
	\begin{subfigure}[b]{\columnwidth}
		\includegraphics[width = 0.9\linewidth]{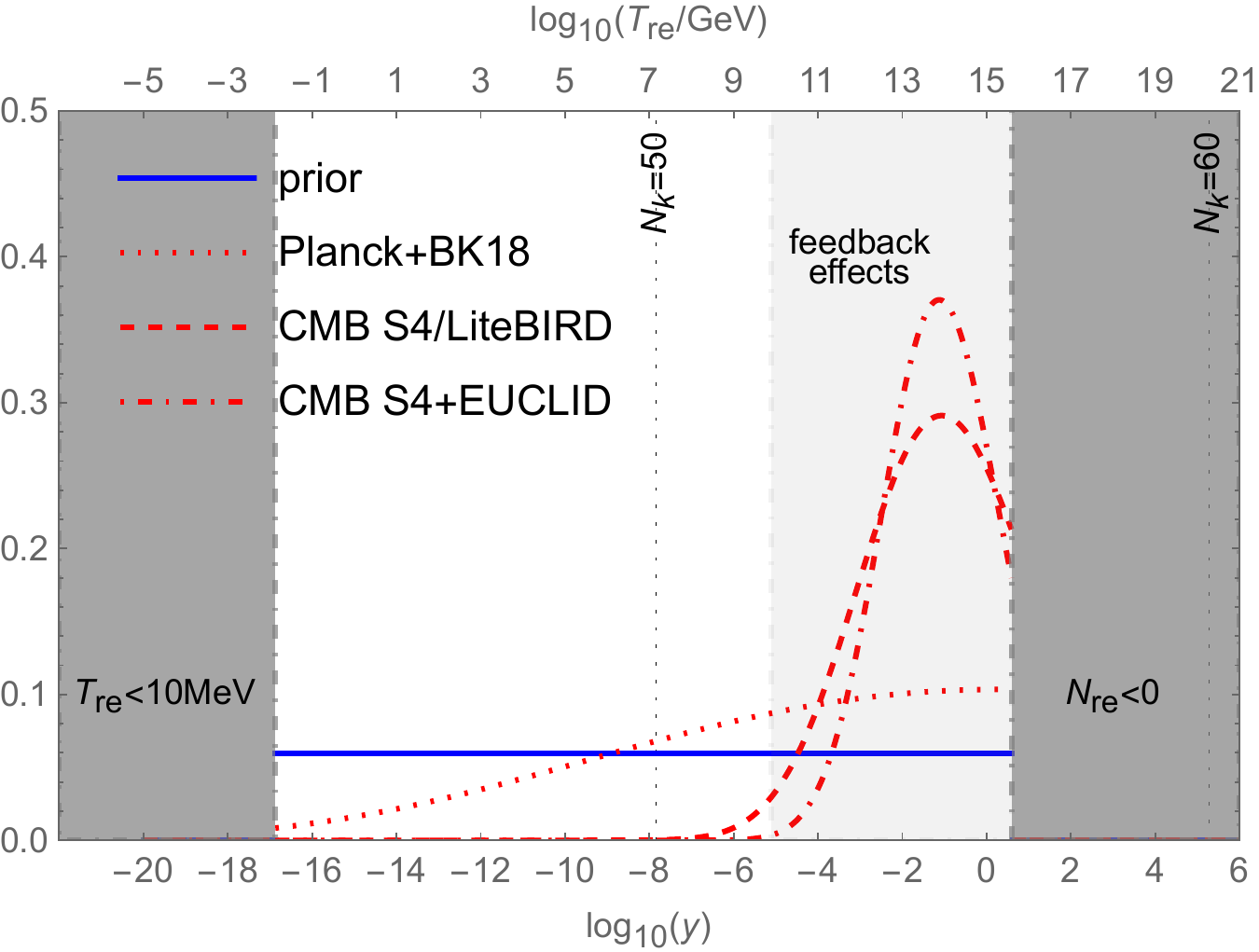}
	\end{subfigure}
	\caption{
	Prior $P(\x)$ and posteriors $P(\x|\mathcal{D})$
	for $\x=\log_{10}\g$
	with the different choices of 
	$\nsbar$,
	$\rbar$, $\sigmans$, and $\sigmar$ in the 
	RGI model (upper panel) and
	$\alpha$-T model (lower panel). 
	We assumed $\gamma=\tilde{\gamma}=1$, but checked that the conclusions remain unchanged when using $\gamma=\Nk'$ or $\tilde{\gamma}=(n_s^{'2} + r^{'2})^{1/2}$, with $'$ indicating a derivative with respect to $\x$.
	\label{Figure2}
	}
	\label{results}
\end{figure}

\paragraph{CMB constraints on the inflaton coupling}
With the above considerations and the relations given in the appendix,  
$n_s$ and $r$ 
in a given model of inflation are simple functions of 
$\x\equiv \log_{10}\g$. 
Prior to any measurement of $(A_s,n_s,r)$
it is known that
$\Nreh > 0$ and that there is a lower bound $\Treh > T_{\rm BBN}$ to allow for successful big bang nucleosynthesis (BBN).
When \eqref{GeneralScaling} and \eqref{GeneralScalingSelf} are fulfilled one can use the standard estimate $\Treh\simeq \sqrt{\Gamma M_{pl}} \times (90/(\pi^2 g_*))^{1/4}$ to obtain a lower bound on the coupling
$\g > g_*^{1/4} \sqrt{c} T_{\rm BBN}/\sqrt{m_\phi M_{pl}}$,
which for plateau models translates into
$\g > \sqrt{c} (T_{\rm BBN}/M_{pl})\times (\frac{g_*}{A_s r})^{1/4}$
\cite{Drewes:2019rxn}.
Hence, we use the prior probability density function (PDF)
\begin{eqnarray}\label{PxdefNew}
P(\x)  
= C_1 \theta\big(\Treh(\x)-T_{\rm BBN}\big)
\ \gamma(\x) \ \theta\big(\Nreh(\x)\big),
\end{eqnarray}
with $\theta$ the Heaviside function,  
and $\gamma$ a function that allows for a re-weighting of the prior $P(\x)$. 
The constant $C_1$ can be fixed from the requirement $\int d\x P(\x) = 1$.
We now quantify the gain in knowledge about $\x$ that can be obtained from data $\mathcal{D}$. 
This gain can be quantified by the posterior distribution $P(\x|\mathcal{D})=P(\mathcal{D}|\x)P(\x)/P(\mathcal{D})$ with $P(\mathcal{D})=\int d\x P(\mathcal{D}|\x)P(\x)$. 
Current constraints from the data $\mathcal{D}$ obtained by Planck and BICEP/Keck \cite{BICEP:2021xfz} 
 can be approximated by the
likelihood function
\begin{eqnarray}\label{Eq:Likelihood}
P(\mathcal{D}|\x) = C_2\mathcal{N}(n_s,r|\nsbar,\sigmans;\rbar,\sigmar)\theta(r)
\tilde{\gamma}(\x),
\end{eqnarray}
with 
$\mathcal{N}(n_s,r|\nsbar,\sigmans;\rbar,\sigmar)$
a two-dimensional Gaussian\footnote{
The parametric dependencies presented in the appendix imply that using the full information about the data made public at \href{http://bicepkeck.org/}{http://bicepkeck.org/} is very unlikely to change our conclusions.
}  
and $\tilde{\gamma}$ another weighting function. The constant $C_2$ is fixed by normalising $P(\mathcal{D}|\x)$ to unity. 
We fix the fiducial
values to $\nsbar = 0.967$, $\rbar = 0.01$ and 
estimate the errors $\sigmans = 0.005$, $\sigmar = 0.018$ based on Fig.~5 in \cite{BICEP:2021xfz}. 
The result is shown in Fig.~\ref{Figure2}.
While it is known that present data already provides information about the reheating epoch \cite{Martin:2014nya}, current CMB observations do not provide a significant information gain on $\x$ with respect to $P(\x)$.

The scalar-to-tensor ratio will be constrained with much higher accuracy in the future \cite{Kamionkowski:2015yta}. 
 To quantify the expected information gain on $\x$ we repeat the analysis for $\rbar=0.02$
 with $\sigma_{n_s}=0.002$  and $\sigma_{r}=0.0012$,
 which reflects the sensitivity anticipated by LiteBIRD \cite{LiteBIRD:2022cnt}  or CMB-S4 \cite{CMB-S4:2020lpa}.\footnote{
Due to the different designs, sky coverages and foregrounds affecting the ground-based CMB-S4 program and the LiteBIRD satellite estimates of  
$\sigma_{r}$ as a function of $\rbar$ vary for both observatories. The values used here, which are based on Fig.~8 in \cite{CMB-S4:2020lpa} and is roughly consistent with Fig.~44 in \cite{LiteBIRD:2022cnt}, are sufficiently accurate for our proof of principle. A more detailed discussion can be found in \cite{Drewes:2023bbs}.
 }
Fig.~\ref{Figure2} shows that in both models future data  can rule out previously allowed values of $\Treh$. 
 In the $\alpha{\rm-T}$ model the posterior peaks in a region where condition \eqref{GeneralScaling} is violated, implying that $\Gamma$ depends on a potentially large number of model parameters $\{\SM_i\}$, 
 and it is impossible to translate a constraint on $\Nreh$ into a model-independent constraint on $\g$. 
 This is a result of the fact that the currently allowed region in Fig.~\ref{Figure1} is very close to the $\Nreh=0$ line.
 One can nevertheless obtain constraints
 $\log_{10}\left(\Treh /{\rm GeV}\right) = 13.1\pm1.4$  and $0.00515 < \M/M_{pl} < 0.00526$ (the latter from \eqref{H_k}).
 In the RGI model, on the other hand, the posterior peaks in a region where condition \eqref{GeneralScaling} is fulfilled, so that future CMB data will permit measuring $\g$ independently of the $\{\SM_i\}$. For the fiducial parameters chosen here, the mean values and variances for the posteriors read $\log_{10}y=-6.5\pm 2.2$, $\log_{10}\left(\Treh /\rm{GeV}\right) = 8.4\pm2.1$ and $\M/M_{pl}=0.00529\pm0.00007$.
 Finally, we estimate the improvement that can be made with data from the EUCLID satellite \cite{EUCLID:2011zbd} and Square Kilometre Array (SKA) \cite{Maartens:2015mra}
by using $\sigma_{n_s} = 0.00085$ \cite{Sprenger:2018tdb}. 
 The resulting posteriors in Fig.~\ref{Figure2} for the chosen values of $\nsbar$ and $\rbar$ give $\log_{10}\left(\Treh /\rm{GeV}\right) = 13.5\pm1.1$ in the $\alpha$-T model and $\log_{10}\left(\Treh/\rm{GeV}\right) = 8.3\pm1.4$ in the RGI model.
 The latter corresponds to $\log_{10}y = -6.6\pm 1.4$.

 \paragraph{Conclusions}
 We introduced a simple analytic method to 
quantify the information gain on the inflaton coupling $\g$
 and the reheating temperature $\Treh$
 from observational constraints on $n_s$ and $r$. 
 When applying it to future CMB observations with LiteBIRD and CMB-S4 we showed for the first time that these missions will be capable of performing the first ever measurement of $\Treh$ in both models considered here. For the chosen fiducial values this can directly be translated into a measurement of $\g$ in the RGI model, while 
 in the $\alpha$-T model
 such a translation would require a specification of further parameters $\{\SM_i\}$.
 Adding information from optical and 21cm surveys can 
further reduce the error bar on $\g$, 
and may help to constrain $\alpha$ and $\g$ simultaneously from data by including observational knowledge on quantities not considered here, such as non-Gaussianities or the running of $n_s$. 
 The inflaton coupling $\g$ did not only crucially shape the evolution of the observable universe through its impact on $\Treh$, but it is also a key parameter that connects models of inflation to theories of particle physics.  
 Measuring this microphysical parameter, even with large error bars, will 
 open up a new window to probe the connection between cosmology and fundamental physics. 
 Hence, our findings add a qualitatively new dimension to the physics cases of future observatories. 

\paragraph{Acknowledgements}
We would like to thank 
Marcos Garcia,
Jan Hamann,
Jin U Kang, 
Eiichiro Komatsu, 
Isabel Oldengott, 
Christophe Ringeval, 
Evangelos Sfakianakis, 
and Yvonne Wong for helpful discussions.
We also thank Bj\"orn Garbrecht, Andrea Giammanco, Jan Hamann and Fabio Maltoni for feedback on the revised version of this manuscript. 
L.M.~acknowledges the State Scholarship Fund managed by the China Scholarship Council (CSC) and the Project funded by China Postdoctoral Science Foundation (2022M723677).
MaD would like to thank the University of New South Wales for their hospitality under the Gordon Godfrey visitor program during part of the work on this project.

\paragraph{Appendix: Relation to observables}

In this appendix we give the relations between the RHS of \eqref{GammaConstraint} and observables. 
A detailed derivation can be found in \cite{Ueno:2016dim} and has been adapted to our notation in \cite{Drewes:2017fmn}.
$\Nreh$ can be obtained from
\begin{eqnarray}
\label{Nre}
    N_{\rm re}&=&\frac{4}{3\Bar{w}_{\rm re}-1}\bigg[\Nk+\ln\left(\frac{k}{a_0 T_0}\right)+\frac{1}{4}\ln\left(\frac{40}{\pi^2g_*}\right)\\
    &&+\frac{1}{3}\ln\left(\frac{11g_{s*}}{43}\right)-\frac{1}{2}\ln\left(\frac{\pi^2M^2_{ pl}r A_s}{2\sqrt{\Vend}}\right)\bigg],\nonumber
\end{eqnarray}
with 
\modified{$g_{s*} \approx g_*$},
$a_0$ and $T_0=2.725~{\rm K}$ the scale factor and the temperature of the CMB at the present time, respectively,
and 
\begin{equation}\label{Nk}
    \Nk=\ln\left(\frac{a_{\rm end}}{a_k}\right)=\int_{\varphi_k}^{\varphi_{\rm end}}\frac{H d\varphi}{\dot{\varphi}}
    \approx\frac{1}{M^2_{pl}}\int_{\varphi_{\rm end}}^{\varphi_k}d\varphi\frac{\V }{\partial_\varphi \V }~.
\end{equation}
The subscript notation 
$H_k, \varphi_k$ etc.~indicates the value of the quantities  
$H, \varphi$ etc.~at the moment when a pivot-scale $k$ crosses the horizon.
$\varphi_k$ can be expressed in terms of 
$n_s$ and $r$
by solving the relations
\begin{equation}
    n_s=1-6\epsilon_k+2\eta_k~, \quad r=16\epsilon_k
\label{nANDr}    
\end{equation}
with the slow roll parameters 
$\epsilon=\left(\partial_\varphi \V/\V\right)^2 M^2_{pl}/2$ and 
$\eta=M^2_{pl}\partial^2_\varphi \V/\V$.
In the slow roll regime, we find 
\begin{equation}
\label{H_k}
	H^2_k=\frac{\V(\varphi_k)}{3M_{pl}^2}~
		=\pi^2 M_{pl}^2\frac{r A_s}{2}.
\end{equation}
with $A_s=10^{-10}e^{3.043}$ \cite{Planck:2018vyg}. 
$\Treh$ can be expressed in terms of the observables $(n_s, A_s,r)$
by plugging \eqref{Nre} with \eqref{Nk} into \eqref{Tre}, 
 $\varphi_k$ is found by solving \eqref{nANDr} for $\varphi_k$, and $\Vend$, and $\varphi_{\rm end}$ can be determined by solving $\epsilon=1$ for $\varphi$.
From \eqref{nANDr} we obtain
\begin{equation}
    \epsilon_k=\frac{r}{16}~,\quad \eta_k=\frac{n_s-1+3r/8}{2},
\end{equation}
from which we find
\begin{eqnarray}\label{TakaTukaUltras}
&&\frac{\partial_\varphi \V(\varphi)}{\V(\varphi)}\Bigl|_{\varphi_k}=\sqrt{\frac{r}{8M_{pl}^2}}, \
\frac{\partial^2_\varphi \V(\varphi)}{\V(\varphi)}\Bigl|_{\varphi_k}=\frac{n_s-1+3r/8}{2M_{pl}^2}\quad
\end{eqnarray}
by using the definitions of $\epsilon$ and $\eta$.
Together with \eqref{H_k} this provides three equations that can be used to relate the effective potential and its derivatives to the observables $(n_s,A_s,r)$. That is sufficient to express $\wrehbar$ and $\Nreh$ in \eqref{Nre} in terms of observables, which is all that is needed to determine the RHS of \eqref{GammaConstraint}.

\raggedright\bibliography{inspire_bibiliography}

\end{document}